\documentclass{emulateapj}

\singlespace

\newcommand{\WMAP}{\textsl{WMAP}}
\newcommand{\be}{\begin{equation}}
\newcommand{\ee}{\end{equation}}
\newcommand{\ba}{\begin{eqnarray}}
\newcommand{\ea}{\end{eqnarray}}
 \newcommand{\Oxford}{{Astrophysics, University of Oxford, %
            Keble Road, Oxford, OX1 3RH, UK}}
\newcommand{\PrincetonPhysics}{{Dept. of Physics, Jadwin Hall, %
            Princeton University, Princeton, NJ 08544-0708}}

\shortauthors{} 
\shorttitle{8 GHz emission} 
\slugcomment{Submitted to the Astrophysical Journal}

\begin{document}

\title{Evidence for Anomalous Dust-Correlated Emission at 8 GHz}

\author{Michelle Lu\altaffilmark{1}}
\author{Joanna Dunkley\altaffilmark{2}}
\author{Lyman Page\altaffilmark{1}}

\altaffiltext{1}{\PrincetonPhysics}
\altaffiltext{2}{\Oxford}
% ---------

\begin{abstract}
In 1969 Edward Conklin measured the anisotropy in celestial emission at 8 GHz with a resolution of 16.2$^\circ$
and used the data to report a detection of the CMB dipole. Given the paucity of 8 GHz observations over large angular scales and the clear evidence for non-power law Galactic emission near 8 GHz, a new analysis of Conklin's data is informative.  In this paper we compare Conklin's data to that from Haslam et al. (0.4~GHz), Reich and Reich (1.4 GHz), and WMAP (23-94 GHz). We show that the spectral index between Conklin's data and the 23~GHz WMAP data is
$\beta=-1.7\pm0.1$, where we model the emission temperature as $T\propto \nu^\beta$.  Free-free emission has $\beta\approx -2.15$,
synchrotron emission has $\beta\approx -2.7$ to $-3$. Thermal dust emission ($\beta\approx1.7$) is negligible at 8 GHz. 
We conclude that there must be another distinct non-power law component of diffuse foreground emission that emits near 10 GHz, consistent with other observations in this frequency range. 
By comparing to the full complement of data sets, we show that a model with an anomalous emission component, assumed to be spinning dust, is preferred over a model without spinning dust at 5$\sigma$ ($\Delta\chi^2= 31$). However, the source of the new component cannot be determined uniquely. 

\end{abstract}

%\keywords{cosmic microwave background -- cosmology: observations}

\section{Introduction}
\label{sec:intro}

Many observations have shown that there is diffuse dust-correlated emission at frequencies below 40 GHz where the thermal emission by dust grains is usually expected to be negligible.  The effect was first seen at large angular scales in the COBE DMR data by \citet{kogut/etal:1996a}, who attributed the phenomenon to the simple co-location of thermal dust and free-free emission. These
observations were confirmed by ground based measurements at finer angular scales \citep{deOliveira-Costa/etal:1997,leitch/etal:1997}. Soon thereafter, it was realized that the dust-correlated emission might in fact be due
to microwave emission by spinning dust grains \citep{Jones1997, draine/lazarian:1998b}. In other words, where there is more thermally emitting dust radiating at frequencies $\nu>100$~GHz, there is also more spinning dust emitting near $\nu\sim 20$~GHz. Since then, the correlated diffuse emission has been seen at high and low Galactic latitudes 
\citep{deOliveira-Costa1998,  deOliveira-Costa1999, leitch/etal:2000, Mukherjee2001, Hamilton2001, Mukherjee2002, mukherjee/etal:2003, bennett/etal:2003c, Banday2003, Lagache2003, deoliveira-costa/etal:2004, Finkbeiner2004WMAP, Davies2006, Fernandez2006, boughn/pober:2007, Hildebrandt2007, Dobler2008a, Dobler2008b, Dobler2009, Miville2008, Ysard2010, Kogut2011, Gold2011}.  At large angular scales, \WMAP\ \citep{bennett2003Fore} suggested that this correlation could also be explained by variable-index synchrotron emission co-located with dusty star-forming regions. However, alternative and subsequent analyses gave more support to the spinning dust hypothesis and attributed the source of the correlation to a population of small spinning grains \citep{deoliveira-costa/etal:2002, Lagache2003, Ysard2010, Macellari2011}. 

Dust-correlated emission at microwave frequencies has also been observed in  more compact regions (e.g., 
 \citet{Finkbeiner2002, Finkbeiner2004, watson/etal:2005, Casassus2004, Battistelli2006, Casassus2006, Iglesias2006,  Casassus2008, Dickinson2009, Scaife2009, Murphy2010, Tibbs2010, Scaife2010a, Scaife2010b, Dickinson2010, ade2011, Vidal2011, Castellanos2011, Lopez2011}).   Most recently, the Planck satellite has shown in detail the presence of a
non power-law component.  A new component of emission is clearly seen in the Perseus and $\rho$ Ophiuchus regions \citep{ade2011}. The component is well modeled as spinning dust, although the range of gas temperatures is large. 

It has long been realized that large area maps in the 5-20 GHz range would be ideal for separating free-free from variable-index synchrotron emission, spinning dust, or any other emission process. For diffuse sources, COSMOSOMAS \citep{Hildebrandt2007} observed the sky between 11 and 17 GHz and identified a dust-correlated component, as did TENERIFE
\citep{deOliveira-Costa1999} observing at 10 and 15 GHz. The ARCADE 2 experiment \citep{Kogut2011} observed at 3, 8, and 10 GHz and also found evidence in support of a new component.  Here, we show that  Conklin's 1969 data \citep{Conklin1969thesis, Conklin1969Nature} further improve constraints in this frequency range and at large angular scales.
 
Although spinning dust is currently the preferred single hypothesis for explaining the correlation, there are $\sim10$ free parameters in the spinning dust model (e.g., \citet{Ali-Haimoud2009}); and there could be multiple processes, including magnetized dust emission \citep{Draine1999}, at work.

The rest of the paper is outlined as follows. We describe the observations in \S\ref{sec:data}, discuss an interpretation of the observations in \S\ref{sec:interpret}, and conclude in \S\ref{sec:discuss}. 
\vskip 1cm

\section{Conklin's observations}
\label{sec:data}

\begin{figure}
\epsscale{1.0} 
\plotone{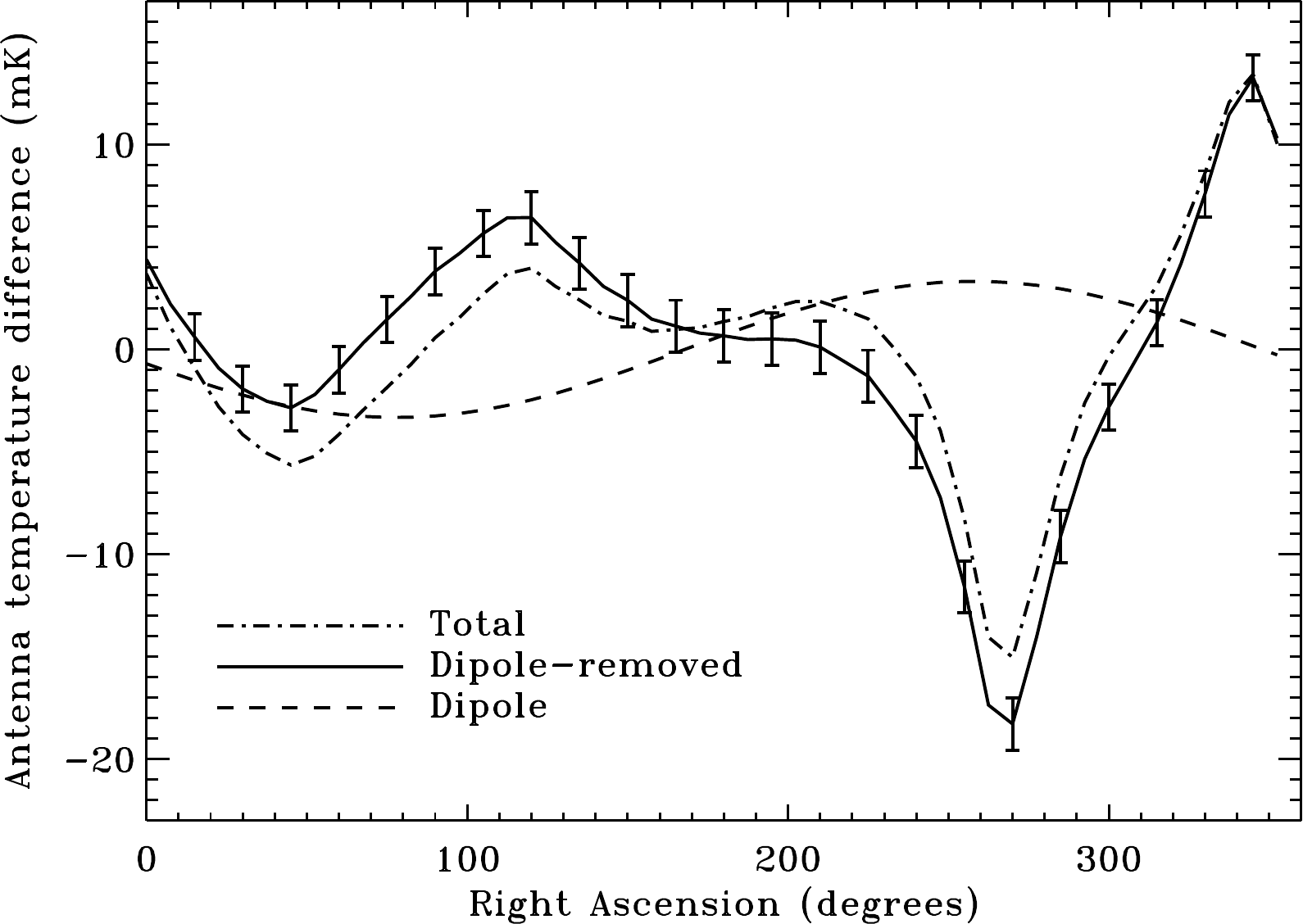}
\caption{Conklin's observations of the anisotropy at 8 GHz. The data are plotted for each 
differential E-W observation at $\delta=32^\circ$, with the now well-established dipole subtracted. The dashed line is the \WMAP\ dipole. The offset is not constrained by the measurement. The peak at RA $\sim300^\circ$ corresponds to  the ``E" beam crossing the Galactic plane at  $b=0^\circ$ and $l=69.5^\circ$.
\label{fig:conklin_data}
}
\end{figure}

Conklin's observations took place on White Mountain in 1968 and 1969, using 
a coherent receiver at 8 GHz with an effective full width at half maximum beam of $\theta_{1/2}=16.2^\circ$. 
The radiometer Dicke-switched at 37 Hz between two $\theta^H_{1/2}=14.5^\circ$ feeds 
pointed $\pm30^\circ$ from the zenith along an E-W baseline. The data were averaged over 4 minutes, the entire apparatus was rotated 180$^\circ$ over one minute, data were averaged again for 4 more minutes, and finally the apparatus was rotated back to its original position. During each such ten-minute cycle, the temperature difference between the two feeds was recorded; Conklin reported the differences as ``east"-``west." The 10-min differences were
then averaged with a 30-min FWHM Gaussian function, effectively broadening the beam profile to $16.2^\circ$ along the scan direction. 
The data we use, shown in Figure~\ref{fig:conklin_data} and reported in Table \ref{table:data}, come from the Nature publication and Conklin's Ph.D. thesis. For this analysis, we use Conklin's raw data after subtracting the contribution from the now well-established dipole. The per-point error bars are deduced from the ``probable error" reported in his thesis.  They may be treated as independent. They are larger than the purely statistical error by a factor of 1.6 {\footnote{To ensure accuracy in calculations, we digitized Conklin's data at twice the resolution reported in Table \ref{table:data}, using 48 bins in RA with error bars increased by $\sqrt{2}$. For the results we report, the calibration uncertainty dominates the errors in the Table.}}.

\begin{figure}
\epsscale{1.15}
\plotone{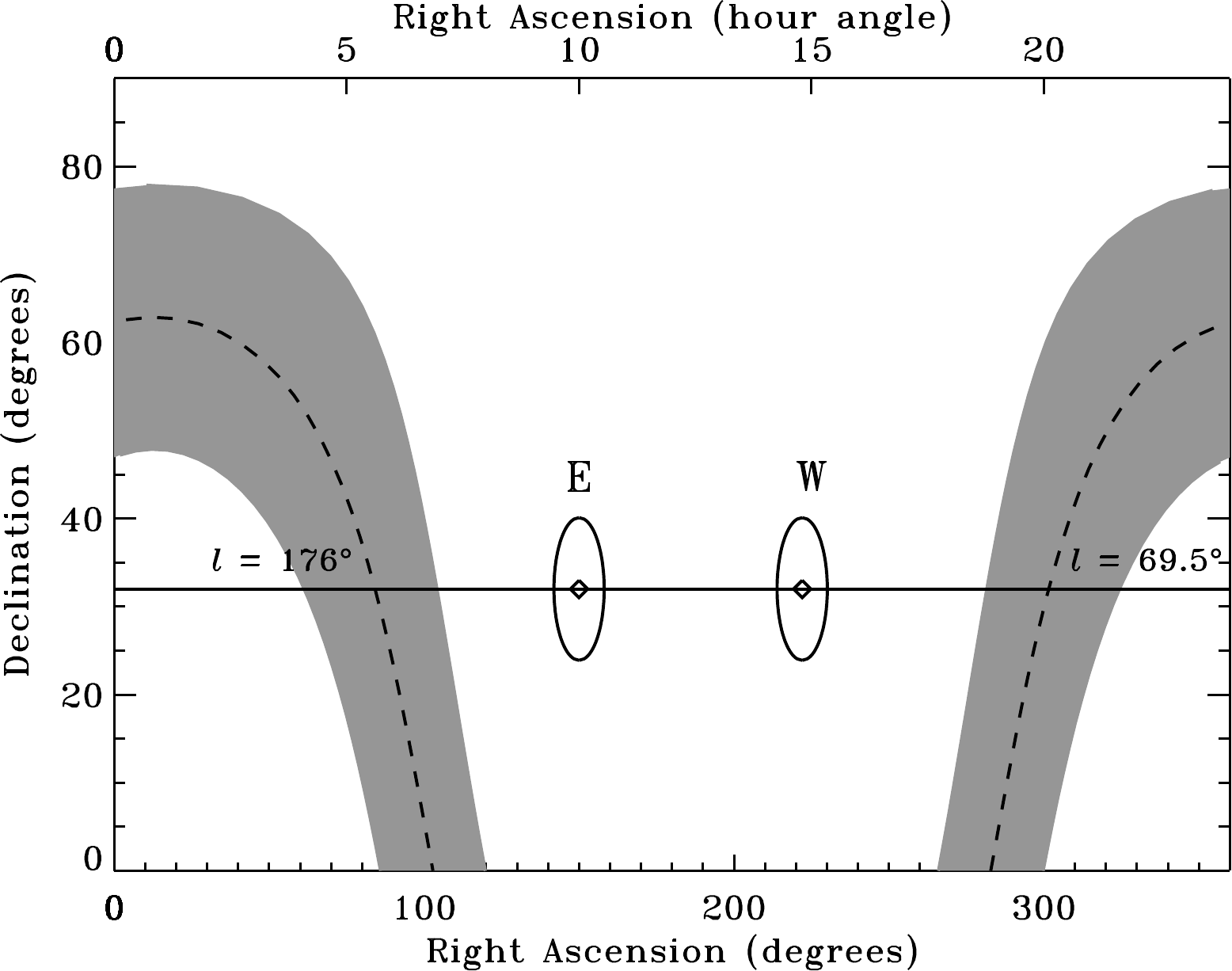}
\caption{Conklin's observations with respect to the Galaxy. The ``E" and ``W" beams are separated by $60^\circ$
and are shown here as circles of diameter $16.2^\circ$. 
The dashed line marks the Galactic plane and the grey swath indicates $b=\pm 15^\circ$. 
\label{fig:conkgal}}
\end{figure}

The measurements were made at a  single celestial circle at $\delta=32^\circ$ as indicated in Figure~\ref{fig:conkgal}. The highest amplitude point at  RA$=300^\circ$ comes from 
when the ``E" beam crosses the Galactic plane at  $b=0^\circ$ and $l=69.5^\circ$, and the ``W" beam is 
out of the plane. The ``E" beam again crosses the Galactic plane at $l=176.4^\circ$. 

After subtracting up to a 10~mK Galactic contribution by 
extrapolating the 404~MHz Pauliny-Toth \& Shakeshaft map \citep{PaulinyToth1962} to 8~GHz, Conklin 
reported a preliminary measurement of the dipole
with amplitude $1.6\pm0.8$~mK in the direction $\alpha=13$h. After Conklin's
1969 Ph.D. thesis, this became $2.3\pm0.9$~mK in the direction $\alpha=11$h
as reported at the IAU 44 symposium \citep{Conklin1972IAU}. Though the error 
bars include an estimate for a $\pm0.1$ uncertainty in the Galactic index 
(the statistical uncertainty was $\approx 0.02~$mK), there were lingering 
doubts about the extrapolation \citep{Webster1974}. Neglecting the correction for the Earth's motion, the \WMAP\ 
dipole ($3.358\pm0.017~$mK in direction $\alpha=11.19$h at $\delta=-6\fdg90$
for the full sky) has amplitude $2.83~$mK at $\delta=32^\circ$,
in excellent agreement. 

\begin{table}
\caption{\small Data at 8GHz from \cite{Conklin1969thesis, Conklin1969Nature}.}
\begin{center}
\begin{tabular}{| l| c |c | }
\hline
RA &	Total 	&Dipole-	\\ 
(deg)&	anisotropy (mK)\tablenotemark{a}	& removed (mK)\tablenotemark{b}		\\ 
\hline \hline
0 &$24.4\pm1.13$&	4.38\\ 
15&$19.8\pm1.13$&	0.60\\ 
30&$16.5\pm1.13$&	-1.94\\ 
45&$15.1\pm1.13$&	-2.86\\ 
60&$16.6\pm1.13$&	-1.00\\ 
75&$18.8\pm1.13$&	1.46\\
90&$21.3\pm1.13$&	3.81\\
105&$23.4\pm1.13$&	5.65\\		
120&$24.7\pm1.27$&	6.42\\
135&$23.1\pm1.27$&	4.20\\		
150&$22.1\pm1.27$&	2.39\\
165&$21.7\pm1.27$&	1.13\\
180&$22.1\pm1.27$&	0.65\\
195&$22.7\pm1.27$&	0.50\\
210&$23.0\pm1.27$&	0.10\\
225&$22.2\pm1.27$&	-1.31\\
240&$19.4\pm1.27$&	-4.50\\
255&$12.4\pm1.27$&	-11.6\\
270&$5.67\pm1.27$&	-18.3\\
285&$14.5\pm1.27$&	-9.13\\
300&$20.4\pm1.13$&	-2.82\\
315&$23.8\pm1.13$&	1.30\\
330&$29.3\pm1.13$&	7.58\\
345&$34.2\pm1.13$&	13.3\\
\hline
\end{tabular}
\tablenotetext{1} {The values include an offset that is removed in our analysis and in Figure 1.}  
\tablenotetext{2} {The errors are the same as for the total anisotropy. Here the offset has been subtracted.}  
\end{center}
\label{table:data}
\end{table}

\subsection{Calibration and beam characteristics}
The system was calibrated with a 1K argon noise tube that
injected power into the feeds through a cross guide coupler. This is effectively a full-beam calibrator.
Conklin was mindful of the required stability and anchored all components with thermal straps 
every two inches. He measured the losses and reflections from 
all components and accounted for the atmosphere. A formal error is not reported
but the system is similar to the one he used for a 10.7 GHz anisotropy measurement \citep{conkbrace1967} where he 
reported $\pm$10\%, which we adopt.

Understanding the beam profile is critical for comparing to other data to assess the Galactic contribution.
During observations, the sky drifts through the H-plane, the profile of which was measured. A measurement of 
the E-plane was not done. Based on his dimensions, we recomputed the full beam pattern using a more precise calculation
\citep{Sletten1988}.
We find an H-plane pattern $\theta^H_{1/2}=14.5^\circ$, in agreement with Conklin, and an E-plane pattern with
$\theta^E_{1/2}=13.7^\circ$ as compared to Conklin's $\theta^E_{1/2}=14^\circ$. For the full beam, we compute a 6\% smaller solid angle.  This translates into a possible bias that the peak-to-peak amplitude of the data is low by 6\%. 
We account for this possibility in the analysis.

\section{Comparison to other data sets}
\label{sec:interpret}

We compare Conklin's data to that from Haslam \citep{Haslam1981, Haslam1982}  at 0.408 GHz; Reich and Reich \citep{Reich1982, Reich1986, Reich2001}\footnote{Obtained from de Oliveira-Costa's Global Sky Model \citep{deOliveira2008}} at 1.42 GHz; \WMAP\ at 23, 33, 44, 60, and 94 GHz \citep{Jarosik2011}; and the FDS map model 8 extrapolated to 94 GHz \citep{Finkbeiner1999}. We adopt calibration errors of 7\%, 4\%, and 2\% for Haslam, Reich  \& Reich, and \WMAP\ respectively. The statistical errors are negligible compared to the uncertainty on Conklin's observations. We convolve the multi-frequency maps to the same 16.2$^\circ$  resolution and extract differenced measurements at $\delta=32^\circ$ to compare to the Conklin data, shown in Figure~\ref{fig:convdiff}. We do not account for the asymmetry in the Conklin effective beam, but test that varying the beam size by $2^\circ$ degrees has a negligible effect on conclusions. In all of our fits
the calibration uncertainties dominate the statistical uncertainties.

\begin{figure}
\epsscale{1.25}
\plotone{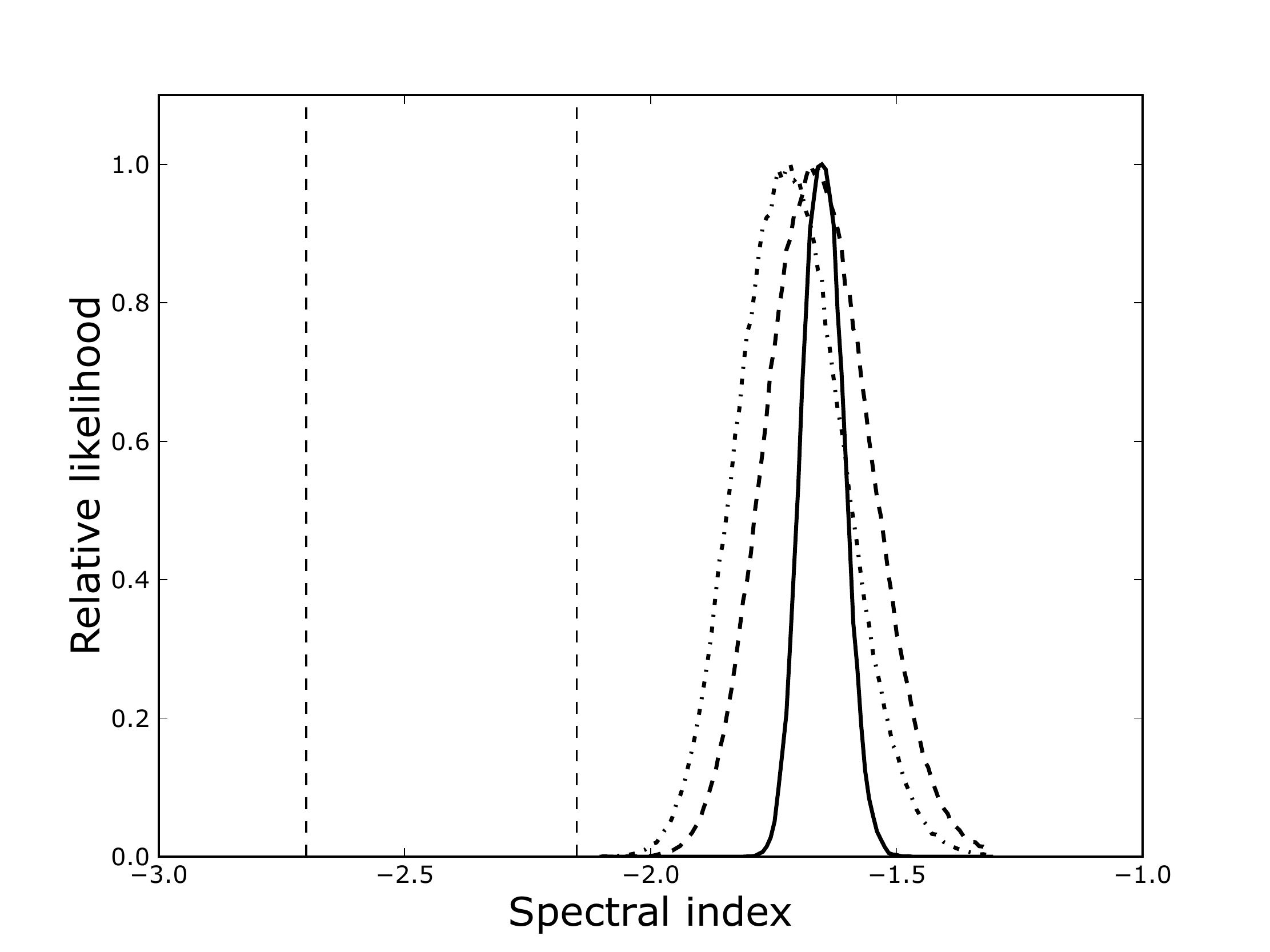}
\caption{Marginalized probability distribution for the effective spectral index between the Conklin 8~GHz data and the \WMAP\ 23~GHz K-band data. The best fit model has $\chi^2 = 64.5$ for 46 degrees of freedom. The index is shallower than would be expected from sychrotron (left vertical dashed line) and free-free emission (right vertical dashed line), indicating the need for an additional component.  The curves are, from left to right, Conklin plus a 6\% systematic bias and calibration error, nominal Conklin with calibration error, and nominal Conklin without calibration error.  The leftmost index shown is $\beta = -1.72\pm0.1$.
\label{fig:diff_index}}
\end{figure}

\begin{figure}
\epsscale{0.95}
\plotone {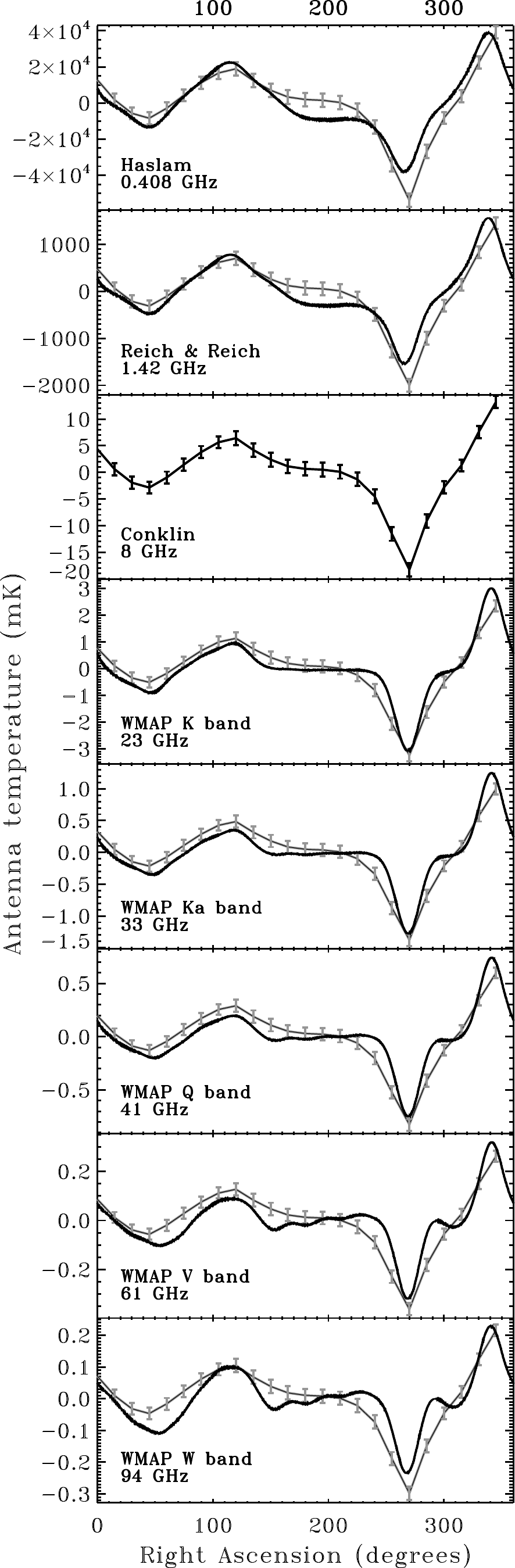}
\caption{Emission intensity at frequencies $0.4<\nu<94$~GHz in the $\delta=32^\circ$ strip surveyed by Conklin compared to the scaled $\nu_0=8$~GHz signal. The data from each map, $I(\nu)$, is fit to the Conklin data using $I(\nu)=s(\nu)I(\nu_0)+A(\nu)$, and compared to the scaled Conklin data using the best-fitting $s(\nu)$ and $A(\nu)$. With multiple emission components, a global scaling of the 8~GHz data does not fully describe the foreground intensity.}
\label{fig:convdiff}
\end{figure}

We initially compare the emission, $I$, at $\nu_0=8$~GHz to the
\WMAP\ K-band emission ($\nu=23$~GHz), using a two parameter model, 
\be
I(\nu_0,\hat{n}) = I(\nu,\hat{n})(\nu_0/\nu)^\beta + A. 
\label{eqn:model}
\ee
We fit for the index, $\beta$, and offset, $A$, finding a
marginalized limit of $\beta=-1.7\pm0.1$, with best-fit 
$\chi^2=64$ (reduced $\chi^2=1.34$). Because \WMAP\ and Conklin are not observing the same mixture of foreground components, the high $\chi^2$ is not surprising.  
Figure~\ref{fig:diff_index} shows the marginalized distribution for the spectral index of the model. We include calibration error and the possible 6\% calibration bias. The index is incompatible with free-free emission ($\beta=-2.15$), synchrotron emission $-3<\beta<-2.7$, or a ``breaking" synchrotron index.
Additionally, the \WMAP\ best-fitting maximum entropy (MEM) model, excluding a spinning dust component, over-predicts the emission at 8 GHz by a factor of two. We conclude that there must exist another component of foreground emission. 

\begin{figure}
{\epsscale{1.15}
\plotone {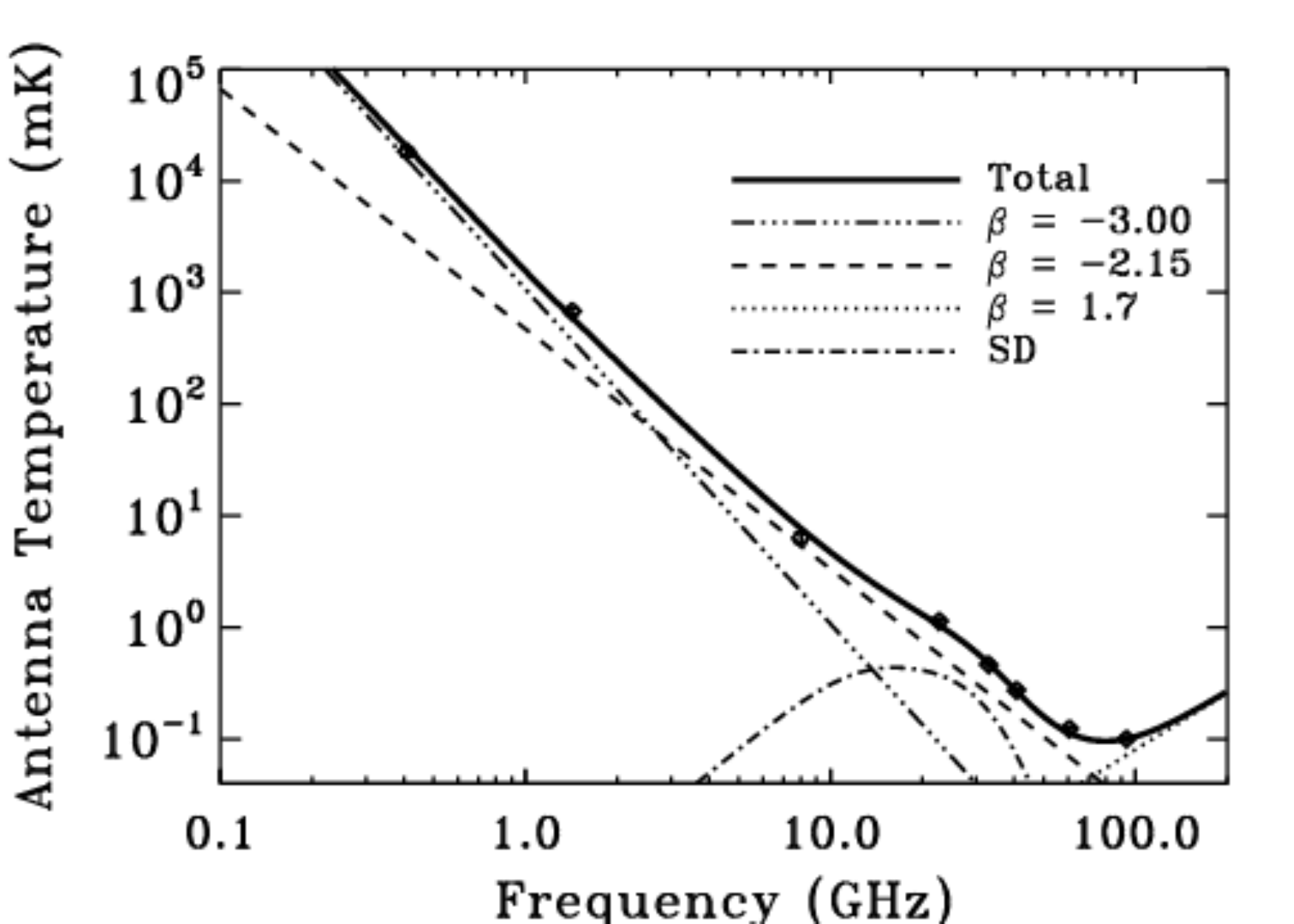}}
\caption{Emission as a function of frequency in the $\delta = 32^\circ$ strip, for observations in the range $0.4<\nu<94$~GHz relative to the 8~GHz observations. The emission is modeled as synchrotron, free-free, thermal dust, and spinning dust, with $\chi^2/{\rm dof}=12.8/3$, compared to $49.9/4$ when excluding spinning dust.
\label{fig:fitfore}}
\end{figure}

To determine the scaling between Conklin's data and each of the data sets in the range $\nu$=$0.4 - 94$~GHz, we fit $I(\nu)=s(\nu)I(\nu_0) + A(\nu)$. Figure~\ref{fig:convdiff}
shows how the intensity measured at each frequency, $I(\nu)$, compared to the scaled Conklin data, $s(\nu)I(\nu_0)+A(\nu)$. The estimated scaling factors are reported in Table \ref{table:scale_fac}, with the goodness of fit. The closest fit is obtained between 8~GHz and the \WMAP\ K band because these are the closest in the logarithm of the observation frequency. A simple power law extrapolation is a poor fit for more widely spaced frequencies as multiple components contribute to the emission. Figure \ref{fig:fitfore} shows the frequency dependence of the estimated scaling factors $s(\nu)$ after multiplying by the {\it rms}
of Conklin's data.  Consistent results are found by fitting to the  $l=176^\circ$ or $l=69.5^\circ$ Galaxy crossings, the {\it rms} of each data set, or to the peak-to-peak amplitude as a function of frequency. The fit is dominated by the Galactic crossing regions.

\begin{table}
\caption{\small Scaling factor between Conklin's data and other data sets.}
\begin{center}
\begin{tabular}{| l  l| c c|}
\hline
Data & Freq (GHz) & $s(\nu)$  & $\chi^2$/DOF\\
\hline
Haslam &0.408 & $2890\pm130$&2.04\\
Reich &1.42  &$107\pm4.54$&1.75\\
Conklin & 8 & 1 & ---\\
WMAP&23 &  $0.179\pm0.0075$&1.33\\
WMAP&33 & $0.0732\pm0.0031$&1.54\\
WMAP&41 & $0.0431\pm0.0018$&1.81\\
WMAP&61 & $0.0194\pm0.00086$&2.69\\
WMAP&94 & $0.0158\pm$0.00072&3.19\\
\hline 

\end{tabular}
\end{center}
\label{table:scale_fac}
\end{table}

To find a physical model that fits Conklin's observations as well as the Haslam, Reich and Reich, and WMAP  data we perform a simultaneous fit to $s(\nu)$ in the range $0.4-94$~GHz. The model includes free-free, synchrotron,
thermal dust, and spinning dust. We test the improvement in $\chi^2$ as we add a spinning dust component, using the model of \citet{Ali-Haimoud2009}.  As free parameters, we take the total hydrogen number density $n_\mathrm{H}$ and the gas temperature $T$, assuming a single density and temperature can be used to model the global emission. For the other parameters, we use the values given by \citet{Ali-Haimoud2009} in their sample spectrum and nominal parameters in \citet{Weingartner2001} Table 1, line 7.  Best fits are obtained with $n_\mathrm{H}$ = 20 cm$^{-3}$ and $T = 300$K. The values of these parameters, however, would be affected by more refined models of the emission process \citep{draine2011}. For the thermal dust emission, we consider two models: one in which the thermal dust amplitude is constrained by the FDS99 model, and one in which the amplitude is allowed to vary (hereafter referred to as ``free dust"). The results are not significantly different,
so we report only the ``free dust" results.\footnote{
For all the models described hereafter, $\nu \equiv \text{frequency}/\text{GHz}$, so that all values are unitless.}

{\it Model 1}.
First, we fit $s(\nu)$  with synchrotron, free-free, and thermal dust emission:
\begin{equation} \label{eq: M1}
M_1(\nu) = a_0\nu^{a_1} + a_2\nu^{-2.15}+ a_3\nu^{1.7}
\end{equation}
where $a_0\nu^{a_1}$ represents synchrotron emission with spectral index $a_1$ and amplitude $a_0$, $a_2\nu^{-2.15}$ represents free-free emission, and $a_3\nu^{1.7}$ represents thermal dust emission with fixed spectral index.  The fit parameters $a_0$, $a_2$, and $a_3$ are required to be positive, and the synchrotron spectral index $a_1$ is varied in the range $-3.5<a_1< -2$. The probability distribution for the four parameters is estimated using Markov Chain Monte Carlo sampling methods. The best-fitting synchrotron index and $\chi^2$ are reported in Table \ref{table:fitfore_chi2}; the model is a poor fit to the data with $\chi^2/\rm{dof}=49.9/4$.

{\it Model 2}.
We then include an additional spinning dust component:
\begin{equation}  \label{eq: M2}
M_2(\nu) = a_0\nu^{a_1} + a_2\nu^{-2.15} +  a_3\nu^{1.7}+ a_4D_s(\nu)
\end{equation}
where $D_s(\nu)$ represents the spinning dust emission template, with amplitude $a_4$. Adding the spinning dust improves the goodness of fit by $\Delta \chi^2=31$ indicating a strong preference for this additional component. 

This model is physically reasonable for a single pixel or direction on the sky. Generally, the relative amplitudes of each emission mechanism spatially varies at any given frequency, and the frequency dependence of the synchrotron, spinning dust, and thermal dust also has spatial variation.  In our case, the model assumes that the relative foreground intensity over the region of sky observed by Conklin may be quantified by a single scaling factor $s(\nu)$.
This is likely over-simplified, but the fit is good.  It is noteworthy that the amplitude of the best fit synchrotron spectrum at \WMAP's 23 GHz band is low compared to the free-free and spinning dust, contributing only $\sim10\%$ of the anisotropy, although the best fit synchrotron spectral index, $\beta\approx-3.00$, is reasonable. One must keep in mind that with so little data the parameter degeneracies are large,  and a more realistic model of the relative contribution of the foreground components would allow for spatially varying amplitude ratios. For example, \citet{Macellari2011} estimate $\sim1/3$ of the all-sky intensity anisotropy to be synchrotron at 23~GHz.

\begin{table*}
\caption{\small Foreground model for total emission along the $\delta=32^\circ$ Conklin strip}
\begin{center}
\begin{tabular}{| l | c | c | c | c |}

	\hline
&	\multicolumn{2} {c|}{Model 1}	& \multicolumn{2}{c|}{Model 2}	\\
&	Nominal	& 	Plus 6\%\tablenotemark{a}		&	Nominal	& 	Plus 6\%\tablenotemark{a} 		\\ 
\hline 
	$a_1$\tablenotemark{b}	&	$-3.23\pm0.12$		&	$ -3.20\pm0.12$		&	$-3.00\pm0.12$	&	$ -2.99\pm0.11$	\\ 
	$\chi^2$/dof 				&	47.2/4		& 	39.7/4		&	11.6/3		&	8.8/3	\\ 
	$\Delta\chi^2$				&	\ldots		&	\ldots		&	35.6		&	30.9	\\ 
\hline
\end{tabular}
\tablenotetext{1} {Includes possible 6\% calibration bias.}
  \tablenotetext{2} {Synchrotron spectral index.}  
\end{center}
\label{table:fitfore_chi2}
\end{table*}

\section{Discussion}
\label{sec:discuss}
Anisotropy in the diffuse microwave emission at 8~GHz, originally measured by Conklin in 1969 to estimate the CMB dipole, now sheds light on Galactic emission. The observed signal is consistent with synchrotron, free-free, and spinning dust emission from the Galaxy. In combination with multi-frequency observations, the 8 GHz data strongly disfavor an emission model with no anomalous dust component, assumed to be rapidly rotating PAH grains. The presence of this additional diffuse component is consistent with observations by ARCADE-2, COSMOSOMAS, and the TENERIFE experiment, and with targeted measurements of dusty regions in the Galaxy. Accurate characterization of the Galactic foregrounds in intensity and polarization is important for extracting cosmological information from the CMB, and will be vital for constraining inflationary models via the large-scale polarization signal. Upcoming low-frequency observations, for example from the C-BASS experiment at 5~GHz \citep{King2010},  will shed further light on the diffuse anomalous dust behavior and allow its properties to be better established.

\acknowledgments 
We gratefully acknowledge the support of the U.S. NSF through award PHY-0355328. We would also like to thank 
Ned Conklin, Angelica de Oliveira-Costa, Clive Dickinson, Bruce Draine, Ben Gold, Al Kogut, David Spergel  and Ed Wollack
for their comments on an earlier draft. This research has made 
use of NASA's Astrophysics Data System Bibliographic Services.  
We acknowledge use of the HEALPix package and
Lambda. We thank Angelica de Oliveira-Costa for helpful comments and assistance with the Reich \& Reich data.

\end{document}